\documentclass{emulateapj}
\usepackage{macros}
\usepackage{graphicx}

\shorttitle{Millisecond Radio Imaging}
\shortauthors{Law et al.}

\begin{document}

 \title{Millisecond Imaging of Radio Transients with the Pocket Correlator}

 \author{C. J. Law\altaffilmark{1}}
 \author{G. Jones\altaffilmark{2,3}}
 \author{D. C. Backer\altaffilmark{1}}
 \author{W. C. Barott\altaffilmark{4}}
 \author{G. C. Bower\altaffilmark{1}}
 \author{C. Gutierrez-Kraybill\altaffilmark{1}}
 \author{P. K. G. Williams\altaffilmark{1}}
 \author{D. Werthimer\altaffilmark{5}}
 \altaffiltext{1}{Radio Astronomy Lab, University of California, Berkeley, CA; claw@astro.berkeley.edu}
 \altaffiltext{2}{Jansky Fellow of the National Radio Astronomy Observatory}
 \altaffiltext{3}{Department of Astronomy, Caltech, Pasadena, CA; jones\_gl@caltech.edu}
 \altaffiltext{4}{Embry-Riddle Aeronautical University, Daytona Beach, FL}
 \altaffiltext{5}{Space Sciences Lab, University of California, Berkeley, CA}

 \begin{abstract}
 We demonstrate a signal processing concept for imaging the sky at millisecond rates with radio interferometers. The ``Pocket Correlator'' (PoCo) correlates the signals from multiple elements of a radio interferometer fast enough to image brief, dispersed pulses. By the nature of interferometry, a millisecond correlator functions like a large, single-dish telescope, but with improved survey speed, spatial localization, calibration, and interference rejection. To test the concept, we installed PoCo at the Allen Telescope Array (ATA) to search for dispersed pulses from the Crab pulsar, B0329+54, and M31 using total-power, visibility-based, and image-plane techniques. In 1.7 hours of observing, PoCo detected 191 giant pulses from the Crab pulsar brighter than a typical $5\sigma$\ sensitivity limit of 60 Jy over pulse widths of 3 milliseconds. Roughly 40\% of pulses from pulsar B0329+54 were detected by using novel visibility-based techniques. Observations of M31 constrain the rate of pulses brighter than 190 Jy in a three degree region surrounding the galaxy to $<4.3$\ hr$^{-1}$. We calculate the computational demand of various visibility-based pulse search algorithms and demonstrate how compute clusters can help meet this demand. Larger implementations of the fast imaging concept will conduct blind searches for millisecond pulses in our Galaxy and beyond, providing a valuable probe of the interstellar/intergalactic media, discovering new kinds of radio transients, and localizing them to constrain models of their origin.
 \end{abstract}

 \section{Introduction}
 Radio waves are rich with physical information encoded in their spectral, polarimetric, and temporal changes. Transient radio sources include neutron stars \citep{m06}, black holes \citep{f99}, the Sun \citep{b98}, and planets \citep{z96}. In all of these examples, the temporal variation of radio emission has been critical to understanding the nature of their emission. This success has inspired several blind surveys for transients with radio interferometers in the hopes of detecting rare events \citep{h05,b07,c10,l10}.

 Fast transients, with time scales shorter than 1 s, hold particular promise for discovery. Physically, transients of this duration are extremely energetic, so they can be detected to very large distances \citep{h03}. Large distances imply a large volume over which they may be detected, increasing the chance to see rare phenomena. Several hypothesized classes of transients may be so rare that they are only detectable over cosmological scales \citep{h01}. At millisecond time scales, dispersion by the interstellar and intergalactic media can (and must) be measured. This opens the possibility of using large samples of transients to measure the density of the intergalactic medium in an entirely new way \citep{c03,m03}.

 New understanding of fast transients has generally come from advances in signal processing. Any astrophysical pulse will have millisecond-scale time structure and megahertz-scale frequency structure induced by dispersion in the interstellar medium. The first millisecond pulsar was discovered by pushing analog signal processing to new time scales \citep{b82}. Modern digital signal processing is advancing the field through flexibility and rapid deployment of new concepts \citep{v09,m10}.

 Here we describe an instrument designed to image the radio sky at millisecond rates with the Allen Telescope Array \citep[ATA;][]{w09}. This system, called the Pocket Correlator (PoCo), is conceptually similar to other correlators for radio interferometers, except that it integrates on time scales about 1000 times shorter. As we describe below, this enables highly efficient surveys for fast radio transients. Section \ref{poco} describes how PoCo was implemented at the ATA. We observed the Crab pulsar and B0329+54 to prove the fast imaging concept and then observed M31 to search for the first millisecond radio pulses from a large external galaxy. In \S \ref{analysis} we compare different algorithms for detecting fast transients. In \S \ref{science}, we present science results from PoCo observations, including limits on the rate of millisecond pulses in M31.  Finally, \S \ref{con}, concludes with potential future implementations of the fast imaging concept.

 \section{Fast Imaging with PoCo}
 \label{poco}

 \subsection{Fast Imaging}
 Fast imaging is the use of radio interferometers to image on time scales less than one second. This time scale is arbitrary, but serves as a useful definition because it describes when pulsar pulse profiles and Galactic dispersion are resolved at radio wavelengths. This is a physical domain not traditionally explored by radio imaging.

 Fundamentally, the benefits of fast imaging come from the fact that an interferometer preserves phase information across its aperture. As a result, an interferometer can encompass and extend upon science currently done with single-dish telescopes. Specifically, fast imaging provides:
 \begin{itemize}
  \item Increased survey speed by using small apertures typical of interferometers,
  \item Improved source localization, which is critical for multi-wavelength identification, and
  \item Better interference rejection and calibration by using multiple independent receivers.
 \end{itemize}

 Large, single-dish radio observatories have traditionally been the workhorses of fast time-domain astronomy \citep[e.g.,][]{c06}. Survey speed scales as $N_{\rm{b}} \Omega_{\rm{b}} (A_{\rm{e}}/T_{\rm{sys}})^2$, where $N_{\rm{b}}$\ is the number of beams on the sky, $\Omega_{\rm{b}}$ is the area covered by each beam, $A_{\rm{eff}}$ is the effective area of the telescope, and $T_{\rm{sys}}$ is the system temperature \citep{c08,d10}. For a dish or array of diameter $D$\ with $N_{\rm{a}}$\ elements, this survey speed scales as $N_{\rm{b}} N_{\rm{a}}^2 D^2$. This shows that large telescopes are very efficient at surveying, especially with the advent of multi-pixel feeds \citep{l06}. However, fast imaging with interferometers consisting of small dishes synthesizes of order hundreds of beams simultaneously; this compensates for their small effective area and gives a competitive survey speed at fast sampling rates. For example, a 42-dish fast-imaging system at the ATA would have a survey speed a few hundred times higher than a 19-beam phased array feed at the Green Bank Telescope.

 These improvements come with the challenge of much greater signal processing in the digital domain and massive data volumes. The large data volumes in time-domain astronomy have motivated a two-step ``detect then analyze'' approach; search algorithms produce candidate transients that are later analyzed in detail. A recent demonstration of how to use interferometers to survey for fast transients is V-FASTR, which detected pulses incoherently across the VLBA \citep{w11,t11}. We developed PoCo to demonstrate the signal processing and algorithms needed to coherently detect pulses (i.e., with visibility data).

 \subsection{PoCo}
 Like many correlators in use today, the standard correlator at the ATA was designed to dump visibilities on timescales of seconds. \citet{l09} describe a test of the fast imaging concept applied to pulsar B0329+54 integrating with the standard ATA correlator at 100-ms rates. This test showed that a correlator could be used to image the on-pulse emission of a pulsar, but did not have the temporal resolution or bandwidth needed to measure dispersion. Therefore a small, special-purpose correlator was conceived as a pathfinding instrument towards future fast-imaging instruments.

 \subsubsection{Detailed Instrument Description}
 \label{instrument}
 Table \ref{capabilities} summarizes the capabilities of the Pocket Correlator. The design used for this experiment was composed of a single ROACH board from the CASPER\footnote{See \url{http://casper.berkeley.edu}.} collaboration combined with a pair of quad-input analog-to-digital converters (ADCs) and a host computer to capture the visibility data. Eight input signals were Nyquist sampled to provide 104 MHz of bandwidth. The eight inputs were connected to a single polarization from each of eight ATA antennas and were down-converted and anti-alias filtered for the second Nyquist zone spanning 104--208~MHz.

 \begin{deluxetable}{lcccc}
 \tablecaption{Summary of PoCo Capabilities \label{capabilities}}
 \tablehead{
 \colhead{Property} & \colhead{Value}
 }
 \startdata
 Antennas & 8 \\
 Channels & 64 \\
 Bandwidth & 104 MHz \\
 Shortest int. time & 78.8 $\mu$s
 \enddata
 \end{deluxetable}

 After digitization, a variable length integer delay of up to 256 samples was applied to each input for coarse geometric delay compensation. The signals were then channelized using a 2-tap polyphase filterbank with 64 channels. The raw complex voltages from each filterbank channel were then multiplied by a scaling factor, after which the real and imaginary parts were rounded and truncated to 4 bits. The data were then transposed to time step, antenna, frequency order. A pipelined architecture \citep{p08} X-engine then calculated all of the cross multiplications and internally accumulated the products for 128 time steps, corresponding to an integration time of $128 \times 128 / 208\rm{MHz} \sim 78.8 \mu\rm{s}$. The data were then grouped into packets and transmitted by 10 gigabit Ethernet to a receiving computer. On the receiving computer, the data were then further integrated to reduce the final data rate before storing to disk. 

 Many aspects of the hardware design were motivated by the rapid deployment of the instrument. In particular, the hardware available on a ROACH board should provide sufficient resources to form at least 1024 channels and to provide a longer onboard integration length. The design presented here fit easily within the resources provided by the FPGA and did not require the complexities of off-chip memory.

 \begin{figure}[tbp]
 \includegraphics[width=0.5\textwidth]{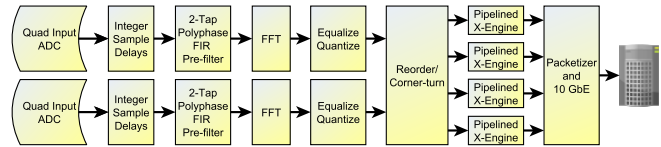}
 \caption{Diagram of the PoCo signal path. \label{BlockDiagram}}
 \end{figure}
 
 \subsection{Observations}
 PoCo was designed and commissioned in the middle of 2010. The instrument was temporarily installed at the ATA, located at the Hat Creek Radio Observatory in northern California. The array is composed of 42 6.1-m dishes, each equipped with a receiver sensitive from 0.5 to 11 GHz \citep{w09}. The entire bandpass was sent as an analog signal over optical fiber into the on-site signal processing room, where PoCo was located.

 Eight, single-polarization, analog antenna signals were sent to PoCo. Only x (horizontal) polarizations were used for all observations described here. Antennas were chosen from the southwest corner of the array, with a maximum baseline of about 170 meters.

 Installation, commissioning, and science observations with PoCo were done from July 23-25, 2010 \footnote{More details available at the observing log at \url{http://log.hcro.org/content/poco}.}. Since PoCo does not correct for geometric delays in real time, the observations were made in short ($\sim1$\ hour) segments during which delay changes were too small to cause decorrelation. In practice, this required calculating the sum of electronic and geometric delays for some time during the observation, fixing those as coarse delays in PoCo, then observing while the delays were valid. A postprocessing script then applied the time-dependent geometric delays (a.k.a. ``fringe rotation'') to each integration, baseline, and frequency channel.

 This method was tested by observing a bright, constant source like Cygnus A over a range of hour angles and measuring the correlation amplitude and delay errors before and after postprocessing. This test showed that the delay errors were stable and small enough to be calibrated. Inspecting the data showed that three antennas had low cross-correlation power, likely due to an error in the PoCo design. Those antennas were flagged and ignored in all subsequent analysis. An additional baseline had cross-correlations with non-astrophysical time and frequency structure, possibly due to analog coupling (``cross talk'') between the signal paths of those two antennas. In the end, PoCo produced nine, science-quality cross-correlated spectra per integration.

 Table \ref{obstable} summarizes the PoCo science observations. To test the system on known fast transients, we observed the Crab pulsar and B0329+54, two pulsars known for having bright individual pulses. The Crab pulsar is located in a bright pulsar wind nebula, while B0329+54 is not detectable outside its pulsed emission. Observing at frequencies from 718 to 822 MHz, a dispersion measure of 57 pc cm$^{-3}$ \citep[typical of the Crab pulsar;][]{m05} produces dispersion smearing up to 2 ms per channel, similar to the typical integration time used. For dispersion measures of 126 pc cm$^{-3}$, the largest used here, the dispersion smearing is at most 5 ms per channel. Observing near 750 MHz balances the need for bright pulses against the sensitivity loss due to dispersion smearing.

 \begin{deluxetable}{lcccc}
 \tablecaption{Summary of PoCo Science Observations \label{obstable}}
 \tablehead{
 \colhead{Target} & \colhead{Mean Freq.} & \colhead{Int. time} & \colhead{Duration} & \colhead{Calibrator} \\
                  & \colhead{(MHz)}        & \colhead{(ms)}      & \colhead{(hour)}   &                     
 }
 \startdata
 Crab pulsar     & 753  & 1.26 & 1.74 & Crab nebula  \\
 B0329+54  & 753  & 1.26 & 0.96 & 3C 147 \\
 M31       & 753  & 1.26 & 0.70 & Cas A 
 \enddata
 \end{deluxetable}

 \subsection{Calibration and Flagging}
 \label{caln}
 The calibration of a fast correlator is largely the same as calibration of traditional correlators. The fact that an interferometer only measures flux on small spatial scales avoids a whole class of systematic errors seen in time-domain studies with single dish observatories \citep[e.g., ground spillover, atmospheric emission;][]{o02}. Cross correlations are also used to calibrate electronic gains of each antenna for accurate absolute flux calibration \citep{b11}.

 For PoCo observations of the Crab pulsar, we use the Crab nebula (a.k.a. Taurus A) for gain and bandpass calibration. The Crab nebula is relatively compact ($\sim$6\arcmin) with a well-calibrated flux \citep[1127 Jy at 770 MHz with a spectral index of --0.3;][]{b77}, making it suitable for self-calibration. The Crab data were gain calibrated at 3 minute intervals, but the solutions show variations smaller than 10\% on hour time scales.

Observations of B0329+54 and M31 could not be self calibrated, so solutions were transferred from observations of 3C~147\footnote{We use the VLA calibration model as implemented in Miriad. See \url{http://www.vla.nrao.edu/astro/calib/manual/polcal.html}} and Cassiopeia A \citep[2417 Jy at 770 MHz with a spectral index of --0.77;][]{b77}, respectively. Because coarse delays were fixed by PoCo, transferring calibration solutions required observing the target/calibrator pairs with the same set of delays. After fringe rotation was applied offline, the difference in geometric delays for the target/calibrator pointings was applied to the calibrator data. This method produced bandpass and gain calibrations appropriate for transferring to the target. The gain amplitude and phase solutions were similar to each other (within 20\%) and the Crab self-calibration solutions, as expected for observations spanning a few hours at the ATA. The B0329+54 data were calibrated from a single solution derived from 3C~147. The M31 data were calibrated from a solution derived from observations of Cassiopeia A made before and after the M31 observation.

 All three gain calibrators have less than a few percent integrated linear polarization at cm wavelengths \citep{a95,w97}. The calibration solutions based on the x-polarization data will not be biased by calibrator polarization. Naturally, a strongly linearly polarized source will not be detected if it is not oriented with the x portion of the feed.

 For simplicity, a single set of flags was applied to all data used here. Roughly half of the band was flagged to remove edge channels and interference. Broad-band interference, possibly from analog television, affected frequencies near 750 and 780 MHz. This flagging left 27 channels for a total bandwidth of 43.9 MHz and mean frequency of 753 MHz. No flagging of individual integrations was done, so intermittent RFI remains in the data. All candidate pulses are visually inspected to reject RFI-triggered detections.

 \subsection{Data Quality}
 \label{quality}
 This section presents visualizations of PoCo complex visibilities to quantify the overall data quality. We used observations pointed at the Crab nebula, 3C~147, and Cassiopeia A to gauge data quality. All of these sources are relatively compact, so their visibilities are easy to interpret. 

\begin{figure}[tbp]
 \includegraphics[width=0.45\textwidth]{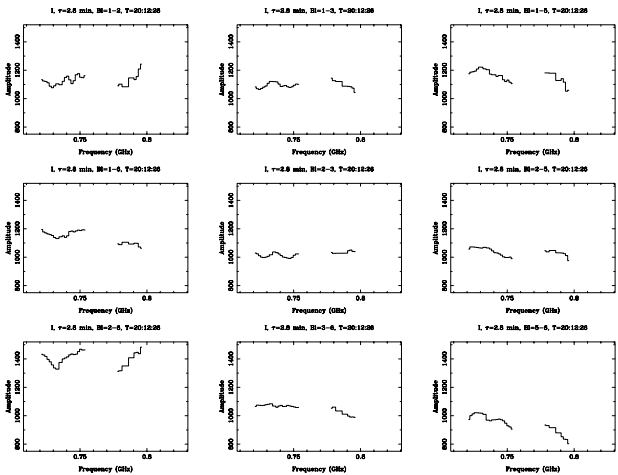}
 \caption{Amplitude spectra in units of Janskys for each of the nine good PoCo baselines toward the Crab nebula. Each plot shows the mean, 64-channel, 104-MHz spectrum averaged over 3 minutes. \label{blspec}}
 \end{figure}

 Figure \ref{blspec} shows calibrated spectra for each baseline during a three minute observation of the Crab nebula. The calibration model (described in detail in \S \ref{caln}) assumes that the Crab nebula is compact, that is, much smaller than the 9\arcmin\ synthesized beam of our subset of the full array. However, after calibrating with this model, there is $\sim$30\% amplitude variation between baselines, with the longer baselines significantly lower than the shortest baselines. This is consistent with a marginally-resolved source with a Gaussian full width at half-max (FWHM) of about 3\arcmin, similar to the size of the Crab nebula \citep{b97}. Since the size of this variation is small (i.e., comparable to the weighting used in imaging algorithms), we keep the simplifying assumption that the nebula is unresolved.

\begin{figure}[tbp]
 \includegraphics[width=0.45\textwidth]{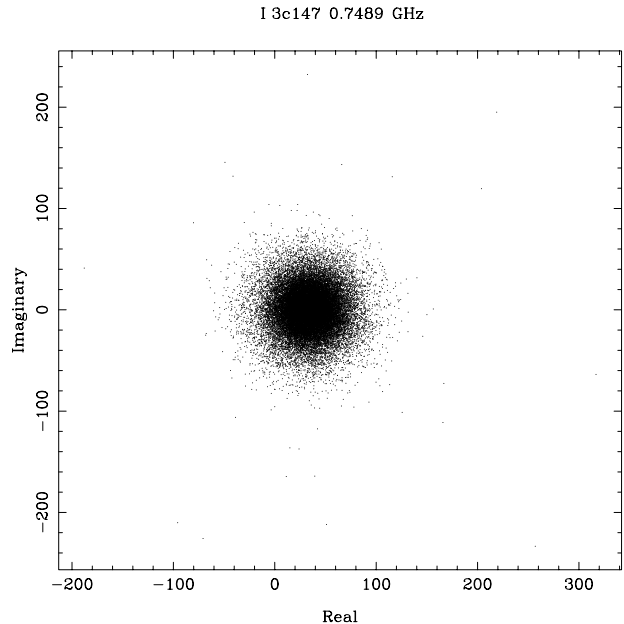}
 \caption{Complex visibilities in units of Janskys averaged over frequency for a 3 minute observation of 3C~147. Each point shows the visibility of a baseline in a 0.77-ms integration in a single channel 1.625 MHz channel. \label{reim}}
 \end{figure}

 Figure \ref{reim} shows the distribution of the visibilities in real-imaginary space for a five minute observation of 3C~147. Each point shows the visibility for a baseline in a 77-ms integration for one channel. The distribution of points in real-imaginary space is predominantly real and Gaussian distributed, showing that there are no significant systematic errors. All targets show similar characteristics down to a level of a $\sim10$\ Janskys on time scales of milliseconds. At longer time scales, all fields show evidence for non-Gaussianity at flux levels of a few Janskys. In all cases, these signals obey closure relations \citep{c89}, indicating that they are associated with celestial emission or calibration errors rather than correlator-based problems.

 The radiometer equation relates the expected sensitivity to observing parameters as
 \begin{equation}
 \sigma_S = \rm{SEFD}/\sqrt{n_{bl} \, \tau \, B},
 \label{rad}
 \end{equation}
 \noindent where SEFD is the system equivalent flux density, $n_{bl}$\ is the number of baselines, $\tau$\ is the integration time, and $B$ is the bandwidth. The distribution of visibilities of 3C~147, the only truly unresolved calibrator observed, can be used to independently measure the noise level. The standard deviation of visibilities per channel, baseline, and 77-ms integration is 24 Jy. Using Equation \ref{rad}, we measure a system equivalent flux density of 8500 Jy. This is consistent with that measured by regular ATA calibration observations and under ideal conditions \citep{w09}. For a single 1.2 ms integration, the PoCo sensitivity is 192 Jy per 1.625-MHz channel and baseline, or 12 Jy per integration using all good baselines and channels. In most cases considered below, the noise level is lower because it is measured in dedispersed data with pulse widths wider than a single integration.

 \section{Analysis}
 \label{analysis}
 While a fast correlator may be thought of as a normal correlator integrating more rapidly, some aspects of transients science at this time scale are new. This section describes dedispersion and source detection with PoCo data, including aspects that are unique to millisecond time scale interferometry. The Miriad software package was used for PoCo calibration and some visualization, while Miriad-Python \footnote{See \url{http://purl.oclc.org/net/pkgwpub/miriad-python}} was used to build the pulse detection and analysis pipelines.

 \subsection{Dedispersion}
 Dispersion introduces an entirely new set of considerations to the analysis of visibility data. In the search for individual pulses, dispersion smears the emission over several integrations and reduces sensitivity \citep{cm03}. 

 For PoCo analysis, we dedisperse by reading blocks of a few thousand integrations (i.e. a few seconds) into Python and calculating dispersion-corrected, complex visibilities for each baseline. The output of the dedispersion process is a single spectrum for each baseline that has been compensated for the time shift for an assumed dispersion measure (DM) and pulse time in the absence of dispersion ($t_0$). The spectrum represents the mean visibility over the pulse duration, which is the quadrature sum of the intrinsic pulse width and the time for the pulse to sweep through each channel.

 The center of the dispersion trial in time is 

 \begin{equation}
 t = 4.15 \times 10^{-3} DM \nu^{-2} + t_0,
 \label{eqn:t}
 \end{equation}

 \noindent where $\nu$\ is defined in GHz, t and $t_0$\ in seconds, and $DM$\ in pc cm$^{-3}$. The pulse width is the time it takes the pulse to disperse through a channel, which is the derivative of Equation \ref{eqn:t}, or

 \begin{equation}
 dt = 8.3 \times 10^{-3} DM \Delta\nu \nu^{-3}
 \end{equation}

 \noindent where $\Delta\nu$\ has units of GHz \citep{cm03}. For pulsar B0329+54 the intrinsic pulse width is 6.6 ms \citep{l95}, while for all other targets, the pulse width is assumed to be much smaller than the 1.2 ms integration time.

\begin{figure}[tbp]
 \includegraphics[width=0.5\textwidth, viewport=0 0 530 380, clip]{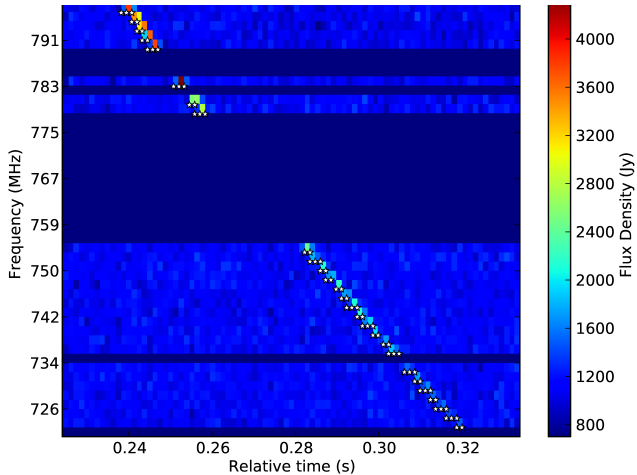}
 \caption{Dynamic spectrum of $\sim800$-Jy pulse seen by PoCo toward the Crab pulsar. The dots show the dispersion track that maximizes the sensitivity to this pulse, assuming a typical Crab pulsar DM of 56.8 pc cm$^{-3}$ \citep{bh07}. Flagged data are blanked and edge channels are not shown. \label{dynsp}}
 \end{figure}

 Figure \ref{dynsp} shows a dynamic spectrum toward the Crab pulsar made by the beamforming technique (see \S \ref{bf}). To our knowledge, this is the first time such a plot has been made from visibility data. The dispersed pulse has a mean flux density of $\sim800$\ Jy and is clearly visible across the PoCo band. The dispersion trial that maximizes the signal-to-noise ratio has $DM=56.8$, consistent with expectations \citep{bh07}; the dispersion track is shown as white dots near the bottom of each channel.

 \subsection{Pulse Detection Algorithms}
 \label{algos}
  This section describes four algorithms for detecting transients in data with dimensions of time, frequency, and spatial frequency. Since an astrophysical signal will be coherent across the array, we show that \emph{imaging} is an efficient way of detecting transients anywhere in the field of view. If the location of the transient is known, an individual pixel can be imaged in a process called \emph{beam forming}. Treating the detection process as an optimization problem, we show that one can detect pulses by \emph{fitting} a model. Finally, we describe a set of \emph{statistical} techniques for detecting pulses. We demonstrate the features and limitations of each algorithm and benchmark their computational demand.

 \subsubsection{Beam forming}
 \label{bf}
 For calibrated visibility data, the real part of the mean cross correlation effectively synthesizes a beam at the array phase center \citep{tms}. Using the real part of the visibility instead of its amplitude has the benefit of being Gaussian distributed with zero mean in the absence of flux. The output of the beam forming process is a value for each channel and integration (e.g., see Figure \ref{dynsp}), reducing the detection problem to that of single-dish transients searches. The sensitivity pattern is the dirty beam centered at the phase center of the array.

 Detecting a pulse in a dynamic spectrum requires searching dedispersed trials for excess power \citep[e.g.,][]{r01}. Because the detection of a pulse is done on the mean of the real part of the visibility, the pulse must be coherent across the array. This coherency is not to be confused with how dispersed pulses are coherently integrated in frequency and time, which is often called ``coherent dedispersion'' \citep[e.g.,][]{h75}. The result of our dedispersion process is a distribution of values in DM-time space that is Gaussian distributed with a mean equal to the brightness at the phase center. Candidate pulses are found by selecting DM trials with brightness greater than 5 times the noise measured in local DM trials. The cumulative probability distribution for a Gaussian function predicts that a 5$\sigma$ threshold will include one false positive per $3.5\times10^6$\ trials; this number is comparable to the number of integrations in an hour of PoCo data, so it is a convenient threshold.

\begin{figure}[tbp]
 \includegraphics[width=0.45\textwidth, viewport=0 0 500 380, clip]{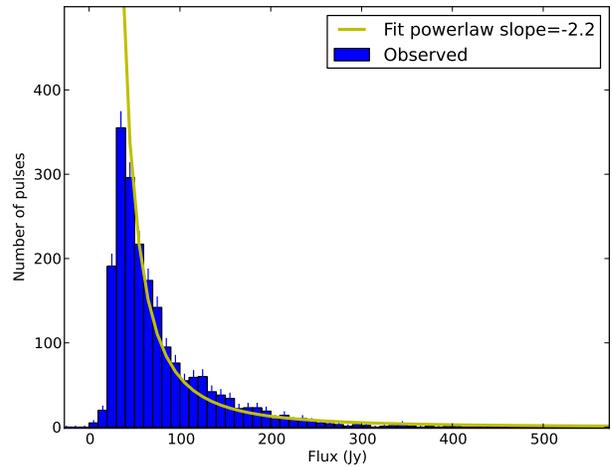}
 \caption{Flux density distribution of B0329+54 pulses detected with greater than 5$\sigma$\ confidence by the beamforming algorithm. The histogram shows the flux density distribution for all 2024 pulses as measured after detection by fitting a powerlaw to their pulse spectrum. Errorbars show the uncertainty in the bin count assuming a Poissonian distribution \citep{g86}. The yellow line shows the best-fit powerlaw to the brighter pulses in the flux density distribution. \label{b0329flux1}}
 \end{figure}

 We demonstrate the beam forming algorithm by searching for pulses in 0.96 hours of data of B0329+54. For a 5$\sigma$\ threshold, the beamforming algorithm detects 2024 candidate pulses, assuming $DM=26.8$ pc cm$^{-3}$ \citep{h04}. This count excludes a few candidate pulses associated with RFI seen in dynamic spectra and 46 candidates detected within the 6.6 ms pulse width (likely multiple detections of the same pulse). After detecting a pulse, a second pass is made to inspect each candidate pulse. Each pulse is fit with a powerlaw to measure its spectral index and mean flux density. The blue histogram in Figure \ref{b0329flux1} shows the distribution of flux density for the 2024 good candidate pulses detected in B0329+54. The flux density is the mean across the assumed pulse width of 6.6 ms and across all good channels, which is approximately equal to the flux density at 753 MHz.

 Figure \ref{b0329flux1} shows the best-fit powerlaw distribution to the pulse brightness distribution. A powerlaw index of --2.2 describes the brightness distribution for all pulses brighter than 40 Jy. Comparing the best-fit powerlaw to the detected pulse distribution shows that the all pulses brighter than approximately 55 Jy are detected. This defines the completeness limit, which is roughly 10 times nominal noise level of 5.7 Jy per DM trial. Table \ref{pulses} summarizes the results of pulse detection algorithms for observations of B0329+54 and the Crab pulsar.

 We detect 1047 pulses above the completeness limit, which is 22\% of the possible pulse detections (i.e., the number of rotations of pulsar during observation). For a mean brightness of $\sim$556 mJy at 753 MHz \citep{h04} and a pulse width of about 6.6 ms \citep[containing 50\% of the pulse flux;][]{l95}, the mean pulse will have a brightness about 30 Jy. The pulse energy distribution of \citet{k03} shows that 10--30\% of pulses should be detected for a completeness limit within a factor of two of the mean brightness. Thus, the detection rate of 22\% for a completeness limit of 55 Jy is consistent with expectations. 

 \begin{deluxetable}{lcccc}
 \tablecaption{Results of Pulse Detection Algorithms \label{pulses}}
 \tablehead{
 \colhead{Source} & \colhead{Algorithm} & \colhead{Total} & \colhead{Completeness} & \colhead{Complete} \\
                  &                     & \colhead{pulses}& \colhead{limit}        & \colhead{pulses}
 }
 \startdata
 B0329+54    &  beamforming   & 2024   & 55 Jy   & 1047 \\
 B0329+54    &  imaging       & 1114   & 65 Jy   & 871 \\
 B0329+54    &  \emph{uv} fit  & 1855   & 55 Jy   & 1043 \\
 Crab pulsar &  beamforming   & 191    & 95 Jy   & 79
 \enddata
 \end{deluxetable}

 Because the beamforming technique is simple and it produces results consistent with expectations, we apply it to science applications with B0329+54 and the Crab pulsar in \S \ref{science}. Below, detections made by other algorithms are compared to the beamforming algorithm to define detection thresholds and other properties.

 \subsubsection{Imaging}
 \label{imgalgo}
 The utility of a fast correlator comes not from beam forming, but from using the visibilities to locate sources. In these kinds of applications, we calculate the mean complex visibility in each channel over a window of width $dt$ centered at $t$. This produces a set of dedispersed spectra for each baseline that can be imaged or fit with a \emph{uv} model.

 A unique aspect of fast imaging is that systematic effects of the array vary much more slowly than the transients. The effects of gain calibration or changes in the \emph{uv}-coverage due to the earth's rotation occur on time scales of minutes to hours. The rapidity of the transients means that visibilities can be differenced to remove systematic effects. In the work presented here, we subtract mean visibilities calculated over a window of 12 ms outside of the DM trial.

 Another unique aspect of fast transients is that they are single, point-like sources. Causality requires a transient of timescale $\Delta t$\ at distance $d$\ to be smaller than $\theta \approx c \, \Delta t / d$. By this reasoning, a millisecond transient at a distance greater than a parsec must be smaller than a micro arcsecond. Further assuming that no more than one fast transient occurs at any moment allows us to simplify our detection algorithm.

 In PoCo analysis, we use these assumptions to build a fast transient detection algorithm based on dirty images. While cleaned images are visually appealing, they are computationally expensive to generate. Our transient detection pipeline instead finds pulses by comparing the peak brightness in dirty images to the median dirty image noise level in other integrations. Since the field likely contains only a point source, the peak brightness of a dirty image is the same as that of a cleaned image. Creating and thresholding a dirty image requires no iterative process (like cleaning or fitting) and uses the Fast-Fourier Transform, so it is relatively computationally efficient. 

 There are many other algorithms for detecting fast transients in this data. One alternative is to image the pulse in individual frequency-time bins. This is contrasted to the technique used here of coherently summing visibilities along a dispersion track before imaging. In analogy with dedispersion in time-domain astronomy, the imaging process is a ``detection'' of the signal, changing the signal from from complex-valued visibilities to real-valued images. As with dedispersion, doing the detection before averaging reduces the sensitivity to the signal by mixing noise with signal in each time-frequency bin.

\begin{figure}[tbp]
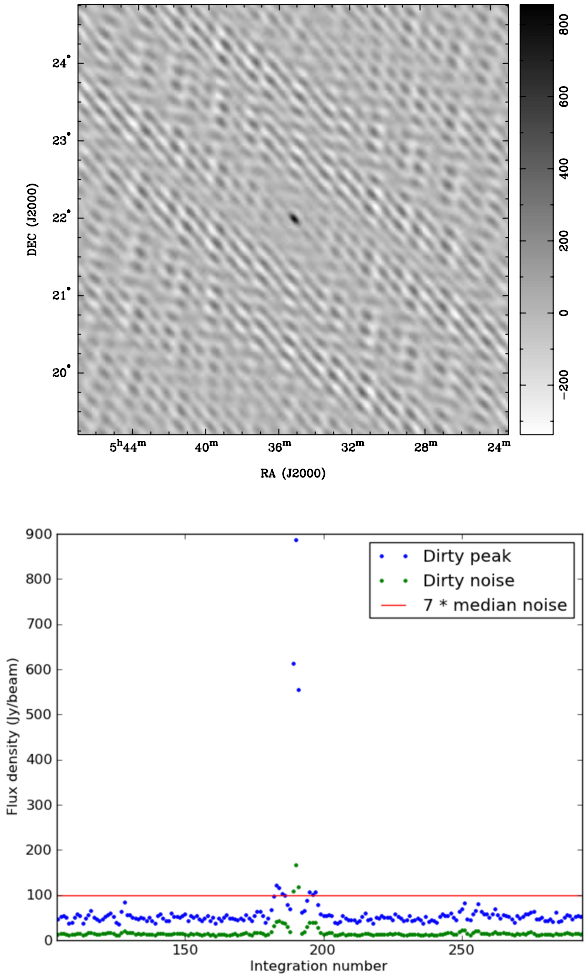

 \begin{center}
 \includegraphics[width=0.4\textwidth]{poco_crab_pulse_image.epsi}

 \includegraphics[width=0.5\textwidth]{poco_crab_pulse_noise.epsi}
 \end{center}
 \caption{(\emph{Top}:) Image of the 1.2-ms, dedispersed pulse shown in Figure \ref{dynsp}. This image shows the peak brightness of the pulse and can also be viewed as a movie. The scale shows the brightness in Jansky beam$^{-1}$. The image and movie are made from background-subtracted visibilities over a window of 10 integrations (12 ms) outside of the pulse. (\emph{Bottom}:) The peak and noise levels in dirty images as a function of time for 200 integrations near the same Crab pulse. Within a window of 10 integrations, the background subtraction produces negative artifacts. \label{image}}
 \end{figure}

 The left side of Figure \ref{image} is a dirty image (also available as a movie at \url{http://www.youtube.com/watch?v=I9U8V-8EaOU}) made from background-subtracted data during the Crab giant pulse shown in Figure \ref{dynsp}. In this and other tests, the dirty images are made using the standard multi-frequency synthesis algorithm with 80\arcsec\ pixels and 250 pixels across to fully cover the primary beam. The right side of Figure \ref{image} shows how the peak and noise level of the dirty images change near the same pulse. The peak pulse brightness is the same in the dirty image as in the cleaned image in the left panel, but the differencing over a 10 integration window produces false detections near the pulse time. We treat all detections within the background subtraction window as a single event.

 The detection threshold for the imaging algorithm is defined to produce the same false-positive rate as the beamforming technique. It is difficult to measure the false-positive rate when only one false positive is expected in the B0329+54 data. A proxy for the false-positive rate is the number of candidate pulses with mean flux density less than the median noise level in a dispersion trial ($\sim5.7$\ Jy). We found that an image peak threshold of 7 times the median noise level produces approximately the number of low-flux candidates as the $5\sigma$\ threshold used with the beamforming algorithm for the B0329+54 data. The false positive rate may be related to non-Gaussianity in the differenced visibilities.

\begin{figure}[tbp]
 \includegraphics[width=0.45\textwidth, viewport=0 0 500 380, clip]{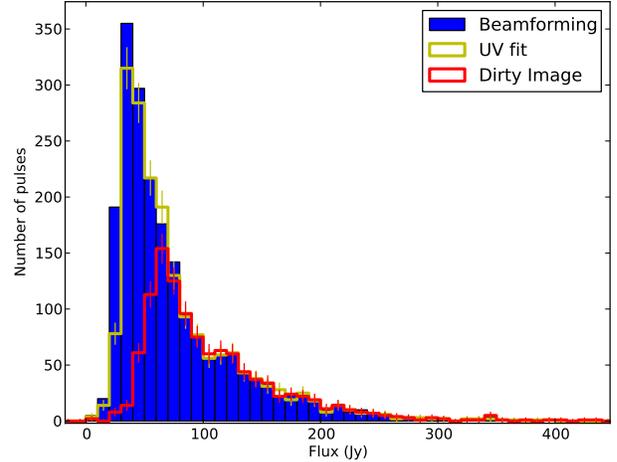}
 \caption{Flux density distribution of pulses detected by three algorithms in 0.96 hours of PoCo data of B0329+54. In all cases, the pulses are first detected, then the beamforming technique is used to generate a pulse spectrum that is then fit with a powerlaw. The light blue bars show a histogram of pulse flux density detected with the beamforming algorithm. The yellow and red lines show the histograms of detections with the \emph{uv}-fit and imaging algorithms, respectively. Errorbars show the uncertainty in the bin count assuming a Poissonian distribution \citep{g86}.  \label{b0329flux2}}
 \end{figure}

 Figure \ref{b0329flux2} shows the pulse mean flux density histogram for candidate pulses detected by the imaging algorithm toward B0329+54. This histogram is compared to the beamforming and \emph{uv}-fit algorithm (described below). In general, the imaging algorithm produced more candidates near RFI, since those images had high peak pixel brightnesses. After removing candidates near RFI seen in the dynamic spectra, we found a set of candidates with negative flux densities located within the background subtraction region of 12 ms; these candidates are also removed. After this filtering, the imaging algorithm detects 1114 unique, good candidates.

 Using the pulse distribution measured by the beamforming algorithm, the imaging algorithm diverges from 100\% completeness at approximately 65 Jy, or 12 times the nominal noise. So the imaging algorithm is approximately 20\% less sensitive than the beamforming algorithm. As summarized in Table \ref{pulses}, 871 unique pulses are detected above the completeness limit in this 0.97 hour observation. More importantly, the histogram from the imaging algorithm is very similar to that of the beamforming algorithm, showing that they are finding the same candidates. Some small differences may arise between algorithms if they associate a given pulse with slightly different integrations.

 As a final test of the imaging algorithm, we simulated observations with pulses away from the phase center. In this case we used data of the Crab with its phase center shifted by 6\arcmin, 24\arcmin, and 1\sdeg. To assure that the gridding of the image didn't affect the signal-to-noise of the detection, the synthesized beam need to be slightly over sampled with an image pixel size 70\arcsec. After making this change, our tests showed that the imaging algorithm detected pulses with identical signal-to-noise ratios regardless of their location in the primary beam.

 \subsubsection{\emph{uv}-fitting}
 \label{uvfit}
 Analogous with the imaging approach to pulse detection, the visibilities can be modeled directly in an approach called ``\emph{uv} fitting''. Fitting visibilities is relatively simple for fast transients applications because the signal is a compact source changing on rapid time scales. The fit also provides robust statistics that can be used to evaluate the likelihood that the candidate is real.

 Our implementation of \emph{uv}-fitting is similar to our imaging algorithms in that DM trials are used to output dispersion-corrected visibility spectra. The \emph{uv} fit is done by running the Miriad task ``uvfit'' on the dedispersed visibility spectra. The task fits a point source sky model to the data and returns best-fit parameter values and their errors. The free parameters of the model are the location of the source and its brightness. 

 As with the imaging algorithm, the threshold for detecting a pulse is set to include the same number of false positives as the beamforming technique. Because the \emph{uv} fit includes the source location, in addition to its peak brightness, the significance of the best-fit peak can be overestimated. Comparing the number of false-positives detected toward B0329+54 using the beamforming algorithm, we set a threshold on the \emph{uv}-fit peak significance of $5.6\sigma$.

 Figure \ref{b0329flux2} shows the mean flux density for all pulses detected by the \emph{uv}-fit algorithm in yellow. As with the imaging algorithm, pulses brighter than the completeness limit have a similar distribution as the beamforming technique. The completeness limit inferred from the powerlaw measured in the beamforming technique approximately 55 Jy and includes 1043 pulses, similar to the beamforming algorithm. Table \ref{pulses} summarizes the results of the \emph{uv}-fitting algorithm in comparison to the other algorithms.

 As with the image fitting technique, we tested how well the \emph{uv}-fitting technique found off-axis pulses. For pulses more than a few beamwidths away from the phase center, the \emph{uv}-fitting technique failed to find the pulse. Experimenting with fitting algorithms showed that the small number of baselines in the PoCo data produced many local minima in the $\chi^2$ surface. This happens for the same reason the synthesized beam has many sidelobes of similar amplitude to the main lobe. In \S \ref{m31}, we demonstrate how iteratively \emph{uv}-fitting throughout the primary beam can be used to test candidate pulses.

 For our goal of finding a pulse-detection algorithm, the \emph{uv}-fitting approach was computationally expensive and had a higher false-positive rate. This is partly because the algorithm was designed to measure properties of the visibilities, which is more than our requirement of a simple detection.

\subsubsection{Statistical}
 As shown in Figure \ref{reim}, all pulse detection processes use the complex valued visibilities to search for deviations from Gaussianity. Since the volume of data to search is large, it is important that the algorithm is as simple and fast as possible. Here we discuss other statistical techniques to detect transients in fast correlator data.

 Generally, an uncalibrated visibility toward a compact celestial source can be written as
 \begin{equation}
 V_i = I \, e^{-j 2\pi (u_i l + v_i m + \phi_i)}
 \label{vi}
 \end{equation}
 \noindent where $u_i$, $v_i$, and $\phi_i$ are the \emph{uv} coordinates and phase error for a given baseline, while $l$\ and $m$\ are the direction cosines of a point source on the celestial sphere \citep{tms}. For a source at the phase center, $l = m = 0$\ and thus $V_i=I e^{-j 2\pi \phi_i}$. For a source detected with an array with many baselines with different phase errors, the visibilities are distributed in an annulus in the real-imaginary plane of radius $I$. Given the structure of Equation \ref{vi}, a compact source anywhere in the field of view of the array will have a similar pattern in the real-imaginary plane.

 The distribution of visibilities in real-imaginary space can be used to detect time dependent astrophysical signals. As noted earlier, differencing of individual visibilities removes steady emission and systematic effects, producing Gaussian-distributed, zero-mean visibilities. This reduces the pulse detection problem to that of rejecting the null hypothesis of Gaussianity, a problem with many potential solutions \citep[e.g., skew and kurtosis; ][]{d71}.

\begin{figure}[tbp]
 \includegraphics[width=0.45\textwidth, viewport=0 0 500 380, clip]{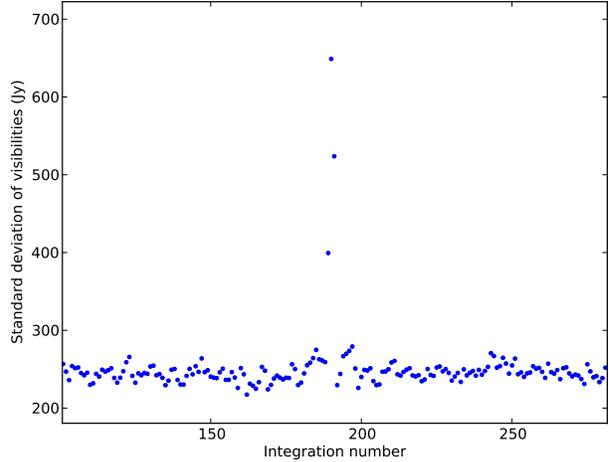}
 \caption{Time dependence of the standard deviation of visibilities in real-imaginary space. This plot shows how the visibilities change during the PoCo observation of the Crab pulse shown in Figures \ref{dynsp} and \ref{image}. \label{crabstd}}
 \end{figure}

 We test this idea using the standard deviation of visibilities in the real-imaginary plane \footnote{The standard deviation of a complex number is equal to the quadrature sum of the standard deviation of the real and imaginary parts.} Figure \ref{crabstd} shows how the standard deviation over all channels, times, and baselines changes during the Crab pulse shown in Figures \ref{dynsp} and \ref{image}. The pulse produces a roughly $28\sigma$\ deviation from the typical value, as compared to a detection significance of $58\sigma$\ for this pulse using the beamforming algorithm. Given the formal connection between gain phase errors and the visibilities of off-axis sources, this technique can also be used to detect pulses in uncalibrated data.

 This statistical test may be thought of as a semi-coherent detection process. While we rely on the coherency of the signal in the visibility domain, we do not use our knowledge of the \emph{uv} coordinates of the baselines. As a result, some information about the signal's coherency is lost. Since this algorithm is not as sensitive as the beamforming or imaging algorithms, we did not apply it beyond this simple test.

 \subsection{Computational Demand}
 A significant challenge for fast imaging is handling the computation required to search for pulses. The computational demand for a pulse search algorithm can be written as a computational demand per search multiplied by the number of searches, or

 \begin{equation}
 C_{\rm{search}} = \mathbb{S} \, \frac{ 2 \, t_{\rm{obs}} }{ t_{\rm{pulse}} } \, \frac{ (\rm{DM}_{\rm{max}} - \rm{DM}_{\rm{min}}) }{ \Delta\rm{DM} }
 \end{equation}

 \noindent where $\mathbb{S}$\ is the per-search demand (measured either in operations or time), $\rm{DM}$\ is dispersion measure, and $t_{\rm{pulse}}$\ is the pulse width. The factor of two comes from Nyquist sampling of the pulse width in time. Rewriting this in terms of observational parameters \citep{cm03}, we find

 \begin{equation}
 C_{\rm{search}} =  \mathbb{S} \, 1.44 \times 10^7 \, t_{\rm{obs}} \, (\rm{DM}_{\rm{max}} - \rm{DM}_{\rm{min}}) \left ( \frac{ \Delta\nu }{ \nu^3 t_{\rm{int}}^2 } \right )
 \label{eqn:c}
 \end{equation}

 \noindent where $\rm{DM}$ is measured in pc cm$^{-3}$, $\Delta\nu$\ is the bandwidth in GHz, $\nu$\ is the observing frequency in GHz, $t_{\rm{int}}$\ is the integration time in milliseconds, and $t_{\rm{obs}}$\ is the observation time in hours. Equation \ref{eqn:c} assumes that pulses are unresolved in time either due to intrinsic or detection effects (e.g., both pulse width and dispersion smearing).

 The details of the pulse search algorithm are contained in the per-search demand factor, $\mathbb{S}$. Thinking of this as a measure of the number of computations, we can quantify scaling relations for our search algorithms. The beamforming algorithm sums visibilities across all baselines and channels for a given dispersion trial, which implies processing demand that scales as $\mathbb{S}_{\rm{bf}} \propto M * N_{\rm{chan}}$. Similarly, any statistical detection algorithm operating on a dispersion trial effectively collapses data along dimensions of channels and baselines and will have a similar computational demand scaling relation. Creating a dirty image requires $\mathbb{S}_{\rm{image}} \propto M + N^2 * log_2(N)$, where $M$\ is the number of baselines and $N$\ is the number of pixels on a side of an image. A \emph{uv} fit is an iterative statistical process, which suggests a scaling like $\mathbb{S}_{\rm{uvfit}} \propto M * N_{\rm{chan}} * N_{\rm{iter}}$, where $N_{\rm{iter}}$\ represents the typical number of iterations required to find a solution.

 The computational demand can also be measured empirically in units of time. We estimated $\mathbb{S}$\ for each algorithm by taking the mean search time over many iterations. These times were measured with a serial version of the pulse search algorithms running on a 2.5 GHz, dual-core Intel core i5 processor. For the beamforming, image, and \emph{uv} fit search algorithms, we found $\mathbb{S}_{\rm{bf}}=0.02$s, $\mathbb{S}_{\rm{image}}=0.28$s, and $\mathbb{S}_{\rm{uvfit}} = 0.10$s, respectively. These relative speeds are consistent with the operations-based scaling relations described above. Note that in this case, the \emph{uv} fit time is underestimated because it searched for pulses near the phase center only; a search over the entire primary beam would inflate the search time by roughly two orders of magnitude.

 The fastest search algorithm described here processes data approximately 20 times slower than the data acquisition rate. Compute clusters will help meet this demand, especially since the search consists of millions of processes that can be run independently. Tree dedispersion algorithms can also improve performance by sharing computations between DM trials \citep{t74}. As described below, a search for pulses in M31 was run with a simple parallelization of running each of 29 different dispersion measure searches on a different processor. This produced an effective dirty image search time of $\mathbb{S}_{\rm{image}}=0.02$s, or 14 times faster than the equivalent series search. Comparing the number of processors and improvement in processing time implies a 50\% parallelization efficiency; more work here will doubtless lead to further improvement.

 \section{Application to Sources}
 \label{science}
 With a working millisecond correlator, we sought to demonstrate the science possible with fast transients. Section \ref{analysis} described algorithms for detecting pulses in fast correlator data. This section describes what happens after detection, when pulses can be studied in more detail. This shows that the science of single-dish time domain astronomy can be implemented for an interferometer.

 \subsection{Crab Giant Pulses}
 \label{crab}
 We used the beamforming technique to search for pulses in 1.74 hours of data of the Crab pulsar. Assuming a DM of 56.8 pc cm$^{-3}$ \citep{bh07}, we detected 191 individual pulses above a $5\sigma$ threshold. The typical noise per dispersion trial was about 12 Jy. This search covered $5\times10^6$\ integrations, from which Gaussian statistics predicts approximately one 5$\sigma$ false positive event. Note that the dispersion track has a typical width of about 2.5 integrations due to dispersion smearing, so the pulse brightness is averaged over 3 ms. This is a factor of 10 larger than the scatter-broadened pulse width of roughly 300 $\mu$s \citep{bh07}, so the detected pulse brightness is 10 times lower than the scatter-broadened pulse brightness.

\begin{figure}[tbp]
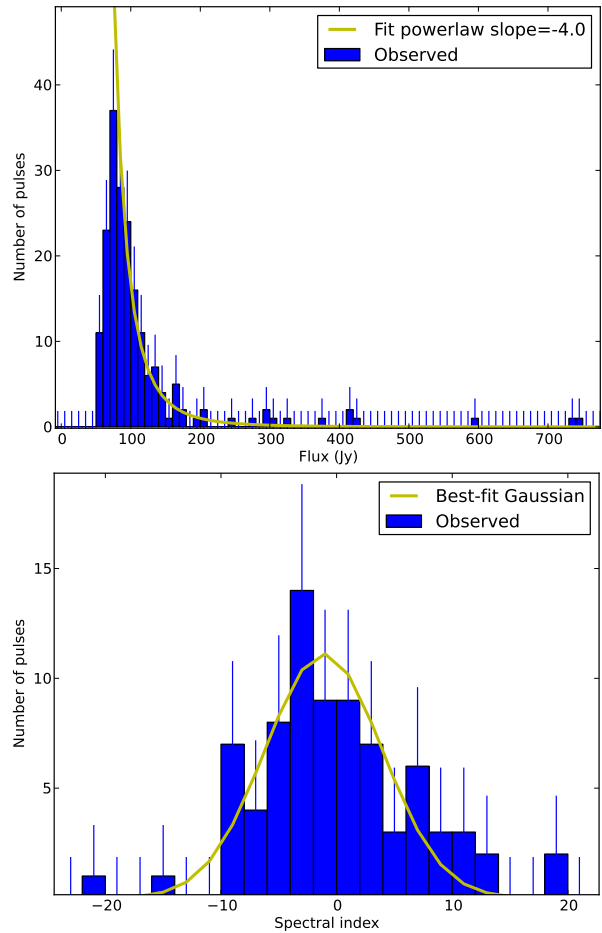

 \begin{center}
 \includegraphics[width=0.45\textwidth, viewport=0 0 500 380, clip]{poco_crab_pulseflux2.epsi}

 \includegraphics[width=0.45\textwidth, viewport=0 0 500 380, clip]{poco_crab_pulseind2.epsi}
 \end{center}
 \caption{\emph{Top:} Brightness distribution of mean Crab pulse flux across PoCo band, centered at 753 MHz. The bars show a histogram of detected pulses brighter than $5\sigma$ and the line shows a powerlaw model fit to all but the lowest two bins. The powerlaw has a slope $\alpha=-4.0$. \emph{Bottom:} A histogram of pulse spectral index fit to all pulses brighter than the completeness limit of 95 Jy. The yellow line shows the best-fit Gaussian with center at $\alpha=-1.1$ and width of 5. \label{crabdisn}}
 \end{figure}

 Figure \ref{crabdisn} shows the Crab pulse brightness and spectral index distributions. The brightness distribution can be fit with a powerlaw distribution of $N\propto S^{\alpha}$. For pulses brighter than 80 Jy, we measure a powerlaw index of $\alpha=-4.0$. This index is similar to the index of --3.5 measured previously for giant pulses \citep{lu95}. The best-fit powerlaw model suggests that the detected pulse distribution is complete above flux densities of 95 Jy, or roughly 12 times the typical noise level of 8 Jy per DM trial. As shown in Table \ref{pulses}, 79 pulses are detected brighter than the completeness limit.

 Considering pulses brighter than the completeness limit, the pulse spectral index distribution is well fit by a Gaussian centered at $\alpha=-1.1$\ with width 5. The width of the spectral index distribution is defined by a combination of sensitivity, bandwidth, and intrinsic effects. It is difficult to separate these effects for pulses over a range of signal-to-noise ratio. However, we note that Crab giant pulses have well-constrained spectral indices throughout this range. For example, the bright pulse shown in Figure \ref{dynsp} has a large, positive index of +19 across the PoCo band.

 \citet{bh07} compile the brightest pulse measured toward the Crab pulsar in 1 hour of observing as a function of frequency. Interpolating to the PoCo observation suggests that we should detect a pulse of brightness several kJy in one hour. Correcting the PoCo apparent pulse brightnesses to the scatter-broadened pulse width, the brightest pulse detected in our observation was 7.5 kJy. The second brightest pulse was 6 kJy. Taking a random 1-hour subset of this observation would likely detect one of these pulses, so the absolute rate of pulses is consistent with the giant pulse rate of the Crab pulsar.

 \subsection{B0329+54}
 As described in \S \ref{bf}, we used the beamforming algorithms to find 2024 pulses above a 5$\sigma$\ threshold, assuming a DM of 26.8 pc cm$^{-3}$ \citep{h04}. The pulse brightness distribution and number of pulses detected is consistent with predictions based on the sensitivity of PoCo.

\begin{figure}[tbp]
 \includegraphics[width=0.45\textwidth, viewport=0 0 500 380, clip]{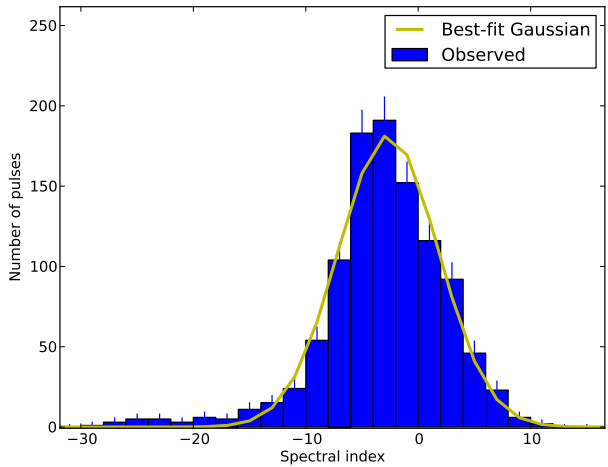}
 \caption{Histogram shows the distribution of spectral indexes fit to B0329+54 pulses brighter than the PoCo detection limit of 55 Jy. The yellow line shows the best-fit Gaussian to the histogram, centered at $\alpha=-2.7$. \label{b0329ind}}
 \end{figure}

 Figure \ref{b0329ind} shows the pulse spectral index distribution for B0329+54 has a roughly Gaussian distribution. The histogram is made from pulses brighter than the completeness limit, which are bright enough to have good spectral fits. The center of the distribution is at $\alpha\approx-2.7$, similar to the mode of the pulse spectral index distribution of --2.5 seen by \citet{k03}. 

\begin{figure}[tbp]
 \includegraphics[width=0.45\textwidth, viewport=0 0 500 350, clip]{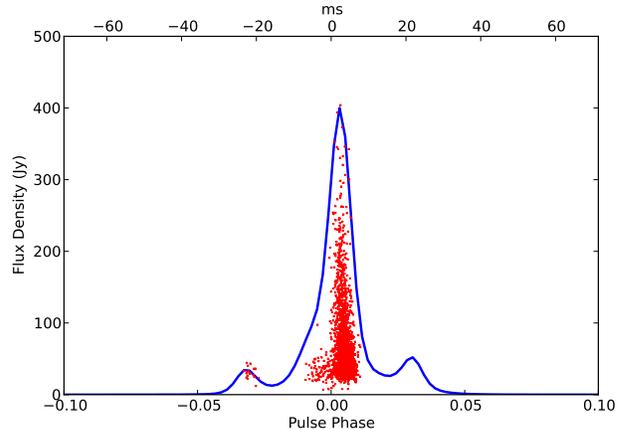}
 \caption{Pulse brightness as a function of pulse phase for PoCo observations of B0329+54. Red dots show the mean pulse brightness of the PoCo pulses near 753 MHz. The blue line shows the 610 MHz mean pulse profile \citep{g98} normalized to match the peak brightness detected by PoCo. \label{b0329ph}}
 \end{figure}

 Figure \ref{b0329ph} shows the brightness of B0329+54 pulses as a function of the pulse phase. The arrival time of PoCo pulses have a bimodal distribution, while the typical pulse profile observed by others has three peaks \citep{g98}. Aligning these two profiles shows that PoCo only detects pulses from the leading and main component, but none from the trailing component. This shows that the PoCo observations coincided with an unusual mode of B0329+54 referred to as ``Abnormal mode B'' \citep{ba82}. This mode occurs approximately 10\% of the time and lasts up to an hour.

 The pulse variability and spectral structure in B0329+54 has been extensively studied as a probe of the interstellar medium \citep{k03}. B0329+54 is in the strong diffractive scintillation regime with $\Delta\nu\approx1$\ MHz and $\Delta t \approx6$\ min. Given our channel size of 1.6 MHz, we expect significant scintillation-induced structure in the pulse spectra that should be stable over a series of about 500 pulses. 

\begin{figure}[tbp]
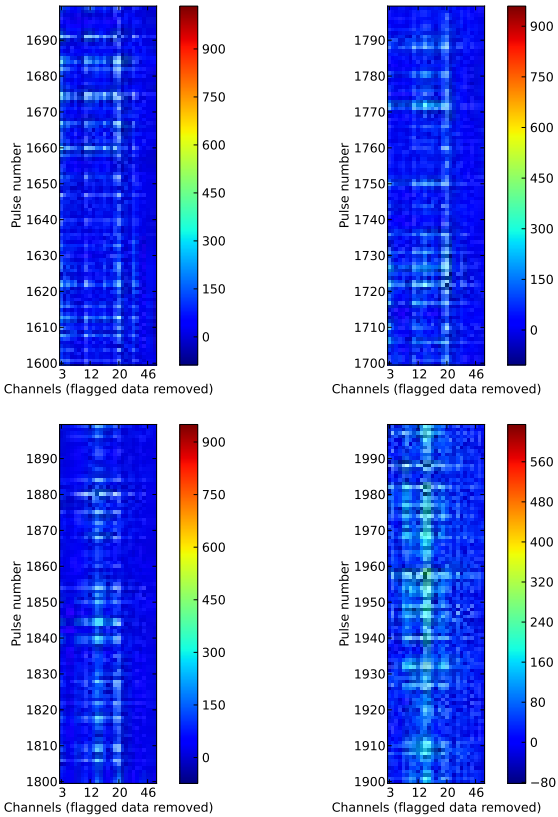

%
 \includegraphics[width=0.23\textwidth,viewport=0 0 300 400, clip]{poco_b0329_173027_fitsp_1600-1700_dynsp.epsi}
 \hfil
 \includegraphics[width=0.23\textwidth,viewport=0 0 300 400, clip]{poco_b0329_173027_fitsp_1700-1800_dynsp.epsi}

 \includegraphics[width=0.23\textwidth,viewport=0 0 300 400, clip]{poco_b0329_173027_fitsp_1800-1900_dynsp.epsi}
 \hfil
 \includegraphics[width=0.23\textwidth,viewport=0 0 300 400, clip]{poco_b0329_173027_fitsp_1900-2000_dynsp.epsi}
 \caption{Dynamic spectra of the pulsed emission from B0329+54. Each panel shows spectra from a sequence of 200 pulses detected by the beamforming algorithm. Brightness in Janskys is coded by a color scale indicated next to the dynamic spectrum. \label{pulsedynsp}}
 \end{figure}

 Figure \ref{pulsedynsp} shows a series of dynamic spectra made of stacked spectra during pulses of B0329+54. Since the PoCo data detect only about 20\% of all pulses, the scintillation timescale of 6 minutes ($\sim500$\ pulses) covers roughly 100 PoCo pulses. Ignoring pulse-to-pulse brightness variation, the PoCo data show that the brightest pulsed emission typically appears in a single channel and that this channel changes on 100-pulse scales. This pattern is consistent with the scintillation time and frequency scales identified previously.

 \subsection{M31}
 \label{m31}
 A major motivation for the development of fast correlators is to enable efficient surveys for extragalactic transients. M31 is a promising source of extragalactic fast transients, since it is the nearest Milky Way-like galaxy and is expected to harbor many Crab-like pulsars \citep{m03}. The ATA field of view at 750 MHz is 5\sdeg, which easily covers the 3\sdeg\ optical disk of M31. This makes M31 a good target to demonstrate the power of fast imaging at the ATA.

 We observed M31 for 0.7 hours using the same band and cadence as in other observations described here (see Table \ref{obstable}). The search for pulses from M31 is more challenging than previous searches in two ways. First, we don't know the location of the pulse, beyond that it should be within M31. This requires the use the dirty imaging pulse search algorithm described in \S \ref{imgalgo}. We assumed that the pulses were temporally unresolved, so background subtraction is expected to reduce sensitivity to pulses longer than a few times the integration time of 1.2 ms. Images were made with 70\arcsec\ pixels and 160 pixels per side, covering more than the 3\sdeg span of the M31 optical disk.

 A second complication to the M31 pulse search is that the pulse dispersion is not known. Following the description of \citet{cm03}, dispersion trials were spaced by 2.7 pc cm$^{-3}$. This gives at least 50\% sensitivity to pulse widths equal to the integration time, assuming a maximum frequency coverage of 73 MHz (the widest separation of unflagged channels). The nominal DM through the Galaxy toward M31 is 67 pc cm$^{-3}$ \citep{c03}. Allowing for uncertainty in the Galactic DM and the potential for DM intrinsic to M31, we search over a DM range from 80\% to 200\% of the Galactic DM to M31. The final search was made with 29 DMs from 50 to 126 pc cm$^{-3}$. 

We searched for pulses using the UC Berkeley compute cluster called Henyey. The search algorithm was run with serial-compiled code, but sending each dispersion measure to different processors in parallel. The data were loaded into the shared memory on each node to reduce I/O latency. The false-positive rate for a 5$\sigma$\ deviation is $2.8\times10^{-7}$. Taking the product of this with the total number of integrations searched, we expect 0.5 false positive per DM trial or a total of about 15 false positives.

 The M31 pulse search produced 92 candidates events. While this is significantly more than the expected false positive rate, our experience with the dirty image algorithm has shown that it is sensitive to RFI. A secondary filter was set by searching data from each candidate with \emph{uv} fit. A compact source was iteratively fit to the data on a grid of fixed locations; the peak brightness was the only free parameter for each fit. The result was a map of the peak brightness throughout the primary beam. We use the peak brightness signal to noise ratio in this map as the final test of the significance of the candidate.

\begin{figure}[tbp]
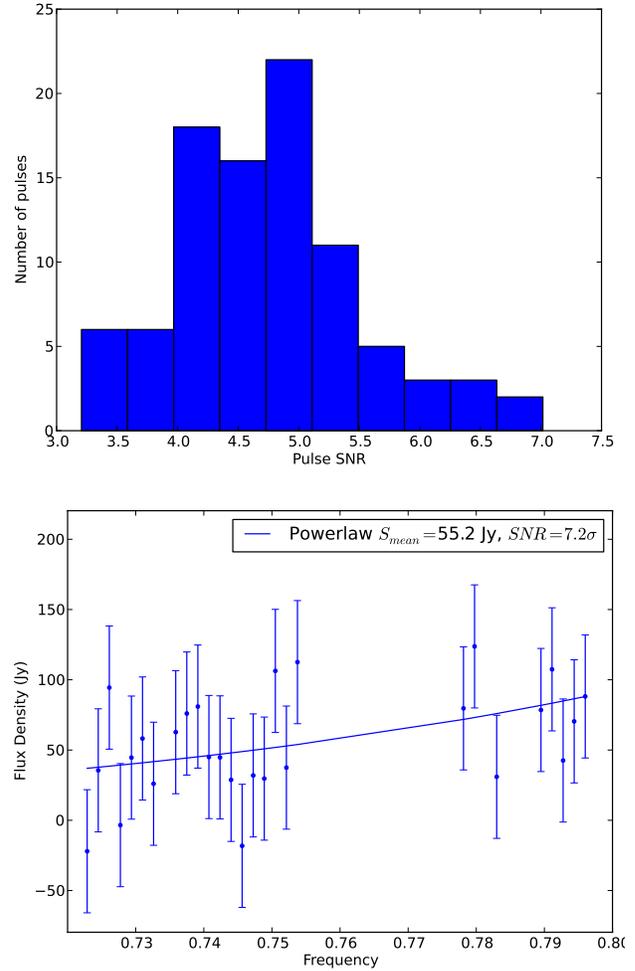

 \includegraphics[width=0.45\textwidth, viewport=0 0 500 380, clip]{poco_m31_cand.epsi}

 \vspace{0.5cm}

 \includegraphics[width=0.45\textwidth, viewport=0 0 500 380, clip]{poco_m31_candspec.epsi}
 \caption{(\emph{Top:}) Histogram of pulse signal to noise ratio of PoCo candidates toward M31. A total of 92 candidate pulses were detected by the imaging algorithm and confirmed with a \emph{uv} fit. The measured signal to noise ratio corresponds to the peak fit across the entire field of view.  (\emph{Bottom:}) Spectrum of the brightest pulse detected by PoCo toward M31. The effective integration time is 3.1 ms. \label{m31cand}}
 \end{figure}

 The left panel of Figure \ref{m31cand} shows a histogram of the peak signal to noise ratio measured by the \emph{uv} fit routine for all 92 candidate pulses in M31. If we assume that set of dirty image candidates is made of true events, normal false-positives, and RFI-induced false positives \emph{and} that the \emph{uv} fit routine is less susceptible to RFI, then thresholding this distribution will select the best events. Under these assumptions, the distribution of candidate pulse brightnesses measured by \emph{uv} fit should be a truncated Gaussian, where events below 5$\sigma$\ are related to RFI and events above 5$\sigma$\ are either true or false positives. This distribution is similar to that seen the left panel of Figure \ref{m31cand}.

 As noted in \S \ref{uvfit}, the \emph{uv} fit algorithm itself has a false positive rate slightly higher than predicted by a Gaussian distribution. Using a threshold of 5.6$\sigma$\ on a \emph{uv} fit search produces the same false-positive rate as a 5$\sigma$\ threshold with the beam forming algorithm. Applying the 5.6$\sigma$\ threshold to the \emph{uv} fit histogram, we find 12 candidates, consistent with the number of false positives expected for the number of dispersion trials. The right side of Figure \ref{m31cand} shows a spectrum measured by beam forming toward the location of the signal to noise peak of the brightest pulse detected by PoCo toward M31. The pulse was detected at a disperison measure of 66.2 pc cm$^{-3}$\ and with lesser confidence in a neighboring dispersion trial. The spectrum was relatively flat, which excludes the possibility of an RFI-triggered event.

 Given that the number of candidates is consistent with the false positive rate for our search, we cannot claim to have a definitive detection of a pulse from M31. Assuming none of these candidate pulses are real, we limit the rate of fast transients in M31. As shown in the Crab observations, the completeness limit is 95 Jy when searching with the exact DM. Our DM spacing allows for a maximum of 50\% signal loss, so our 100\% completeness limit is 190 Jy. Assuming the pulse occurrence probability follows a Poissonian distribution, a nondetection limits the rate of millisecond pulses to less than 4.3 hr$^{-1}$ at 95\% confidence.

 A likely source of fast transients in M31 would be Crab-like pulsars. If Crab-like pulsar was located in M31, could we detect it? In the 1.7 hours of observing the Crab pulsar with PoCo, the brightest pulse was 710 Jy. Scaling this pulse brightness to the distance of M31 \citep[772 kpc;][]{r05} reduces it to a factor of $4\times10^4$ lower than our completeness limit. For a pulse energy powerlaw index of --3.5 \citep{lu95}, a single Crab-like pulsar in M31 is would produce giant pulses detectable at the PoCo completeness limit at a rate of $4\times10^{-17}$\ hr$^{-1}$. While this extremely low rate would be boosted by the fact that PoCo is sensitive to all Crab-like pulsars in M31 \citep{m03}, our nondetection is easily consistent with expectations. Improving sensitivity of fast imaging systems will go a long way to detecting this source population, given its steep powerlaw index. For example, a fast imaging system at the EVLA would have a $5\sigma$\ detection limit of about 25 mJy and could detect a Crab-like pulsar in M31 at a rate of $1\times10^{-3}$\ hr$^{-1}$.

 \section{Conclusions}
 \label{con}
 We have demonstrated a new concept in radio interferometry known as fast imaging. Observations with PoCo at the ATA has shown that the information available to radio interferometers can be used to encompass and extend upon the science of single-dish time domain astronomy. The principle challenges to implementing PoCo were the designing of a correlator to process multiple antennas at millisecond rates and efficiently searching the visibility data. The success of PoCo shows that these challenges can be met today to produce scientifically productive fast imaging systems. 

 PoCo observations of known bright pulsars, B0329+54, and the Crab pulsar, were used to show that the system is sensitive to bright, millisecond pulses. We demonstrate several visibility-based pulse detection techniques and show they can detect and localize pulses anywhere within the field of view of an interferometer. Since visibility data have higher dimensionality than that typically used in transients searches, these algorithms only scratch the surface of potential pulse search techniques. Measuring the computational demand of these search algorithms shows that more work is needed to process data as fast as it is acquired. Compute clusters will go a long way in this effort, since the search algorithms are very easy to parallelize.

 A short PoCo observation of M31 has placed the first constraints on the rate of fast transients from the entire galaxy simultaneously. Our search imaged the entire disk of M31 over a range of DMs with a completeness limit of 190 Jy. No pulses were detected at this level, limiting the rate of millisecond-duration events in M31 to less than 4.3 hr$^{-1}$.
 
 Future developments will expand greatly on the science presented here. A fast imaging system at the EVLA would have the sensitivity to detect RRATs in our Galaxy or giant pulses in nearby galaxies. Scaling from PoCo, a fast imaging system at the EVLA would detect Crab-like pulses from nearby galaxies in observations lasting tens of hours to tens of days. With an efficient pulse-detection algorithm, such a system could conduct real-time searches for pulses commensally with other observations to probe cosmologically significant volumes. Such surveys could find as-yet unknown classes of fast transients, while simultaneously localizing them and measuring their dispersion to constrain models of baryon density in the local universe.

 \acknowledgements{We thank the staff of the Radio Astronomy Lab, SETI Institute, and Hat Creek Radio Observatory for support and Matthew Bailes for comments. GJ is supported by the NRAO, which is operated by Associated Universities, Inc., under cooperative agreement with the National Science Foundation. This research has made use of NASA's Astrophysics Data System Bibliographic Services. 

The authors acknowledge the generous support of the Paul G. Allen Family Foundation that has provided major support for design, construction and operations of the ATA. The U.S. Naval Observatory provided significant funds for ATA construction. Contributions from Nathan Myhrvold, Xilinx Corporation, Sun Microsystems and other private donors have been instrumental in supporting the ATA. The US National Science Foundation grants AST-0321309, AST-0540690 and AST-0838268 have contributed to the ATA project.}


\begin{thebibliography}{}

 \bibitem[Anderson et al.(1995)]{a95} Anderson, M. C., Keohane, J. W., \& Rudnick, L. 1995, ApJ, 441, 300

 \bibitem[Baars et al.(1977)]{b77} Baars, J. W. M., Genzel, R. Pauliny-Toth, I. I. K., \& Witzel, A. 1977, A\&A, 61, 99

 \bibitem[Backer et al.(1982)]{b82} Backer, D. C, Kulkarni, S. R., Heiles, C., Davis, M. M., \& Goss, W. M. 1982, Nature, 300, 615

 \bibitem[Bartel et al.(1982)]{ba82} Bartel, N., Morris, D., Sieber, W., \& Hankins, T. H. 1982, ApJ, 258, 776

 \bibitem[Bastian et al.(1998)]{b98} Bastian, T. S., Benz, A. O., \& Gary, D. E. 1998, ARA\&A, 36, 131

 \bibitem[Barott et al.(2011)]{b11} Barott, W. C., Milgrome, O., Wright, M., MacMahon, D., Kilsdonk, T., Backus, P., \& Dexter, M. 2011, RaSc, 46, 1016

 \bibitem[Bhat et al.(2007)]{bh07} Bhat, N. D. R. et al. 2007, ApJ, 665, 618

 \bibitem[Bietenholtz et al.(1997)]{b97} Bietenholz, M. F., Kassim, N., Frail, D. A., Perley, R. A., Erickson, W. C., \& Hajian, A. R. 1997ApJ, 490, 291

 \bibitem[Bower et al.(2007)]{b07} Bower, G. C., Saul, D., Bloom, J. S., Bolatto, A., Filippenko, A. V., Foley, R. J., \& Perley, D. 2007, ApJ, 666, 346

 \bibitem[Cordes \& Lazio(2003)]{c03} Cordes, J. M. \& Lazio, T. J. W. 2003, astro.ph, 1598

 \bibitem[Cordes \& McLaughlin(2003)]{cm03} Cordes, J. M. \& McLaughlin, M. A. 2003, ApJ, 596, 1142

 \bibitem[Cordes et al.(2006)]{c06} Cordes, J. M. et al. 2006, ApJ, 637, 446

 \bibitem[Cordes(2008)]{c08} Cordes, J. M. 2008, ASPC, 395, 225

 \bibitem[Cornwell(1989)]{c89} Cornwell, T. J. 1989, Sci, 245, 263

 \bibitem[Croft et al.(2010)]{c10} Croft, S. et al. 2010, ApJ, 719, 45

 \bibitem[D'Addario(2010)]{d10} D'Addario, L. 2010, SKA memo series, 123

 \bibitem[D'Agnostino \& Pearson(1971)]{d71} D'Agostino, R. B. \& Pearson, E. S. 1971, Biometrika, 58, 341

 \bibitem[Fender et al.(1999)]{f99} Fender, R. P., et al. 1999, MNRAS, 304, 865

 \bibitem[Gehrels(1986)]{g86} Gehrels, N. 1986, ApJ, 303, 336

 \bibitem[Gould \& Lyne(1998)]{g98} Gould, D. M. \& Lyne, A. G. 1998, MNRAS, 301, 235

 \bibitem[Hankins et al.(2003)]{h03} Hankins, T. H., Kern, J. S., Weatherall, J. C., \& Eilek, J. A. 2003, Nature, 422, 141

 \bibitem[Hankins \& Rickett(1975)]{h75} Hankins, T. H. \& Rickett, B. 1975, Meth. Comp. Phys., 14, 55

 \bibitem[Hansen \& Lyutikov(2001)]{h01} Hansen, B. M. S. \& Lyutikov, M. 2001, MNRAS, 322, 695


 \bibitem[Hobbs et al.(2004)]{h04} Hobbs, G., Lyne, A. G., Kramer, M., Martin, C. E. \& Jordan, C., 2004, MNRAS, 353, 1311-1344

 \bibitem[Hyman et al.(2005)]{h05} Hyman, S. D., Lazio, T. J. W., Kassim, N. E., Ray, P. S., Markwardt, C. B., \& Yusef-Zadeh, F. 2005, Nature, 434, 50


 \bibitem[Kramer et al.(2003)]{k03} Kramer, M., Karastergiou, A., Gupta, Y., Johnston, S., Bhat, N. D. R., \& Lyne, A. G. 2003, A\&A, 407, 655

 \bibitem[Law et al.(2009)]{l09} Law, C. J., MacMahon, D., \& Wright, M. 2009, ATA Memo Series, 85

 \bibitem[Lazio et al.(2010)]{l10} Lazio, T. J. W. et al. 2010, AJ, 140, 1995

 \bibitem[Lorimer et al.(1995)]{l95} Lorimer, D. R., Yates, J. A., Lyne, A. G., \& Gould, D. M. 1995, MNRAS, 273, 411

 \bibitem[Lorimer et al.(2006)]{l06} Lorimer, D. R. et al. 2006, MNRAS, 372, 777

 \bibitem[O'Neil(2002)]{o02} O'Neil, K. 2002, Single-Dish Radio Astronomy: Techniques and Applications, Editors: S. Stanimirovi\'c, D. R. Altschuler, P. F. Goldsmith, and C. J Salter, page 293

 \bibitem[Lundgren et al.(1995)]{lu95} Lundgren, S. C. et al. 1995, ApJ, 453, 433

 \bibitem[Macquart et al.(2010)]{m10} Macquart, J.-P. et al. 2010, PASA, 27, 272

 \bibitem[Manchester et al.(2005)]{m05} Manchester, R. N., Hobbs, G. B., Teoh, A. \& Hobbs, M., AJ, 129, 1993

 \bibitem[McLaughlin \& Cordes(2003)]{m03} McLaughlin, M. A. \& Cordes, J. M. 2003, 596, 982

 \bibitem[McLaughlin et al.(2006)]{m06} McLaughlin, M. A. 2006, Nature, 439, 817

 \bibitem[Parsons et al.(2008)]{p08} Parsons, A. et al, 2008, PASP, 120, 1207

 \bibitem[Ransom(2001)]{r01} Ransom, S. M. 2001, PhD Thesis, Harvard University, 123

 \bibitem[Ribas et al.(2005)]{r05} Ribas, I., Jordi, C., Vilardell, F., Fitzpatrick, E. L., Hilditch, R. W., \& Guinan, E. F. 2005, ApJ, 635, L37

 \bibitem[Thompson et al.(2004)]{tms} Thompson, A. R., Moran, J. M., \& Swenson Jr., G. W. 2004, ``Interferometry and Synthesis in Radio Astronomy'', Wiley-VCH

 \bibitem[Thompson et al.(2011)]{t11} Thompson, D. R., Wagstaff, K. L., Brisken, W., Deller, A. T., Majid, W. A., Tingay, S. J., Wayth, R. B. 2011, astro-ph/1104.4900, accepted to ApJ

 \bibitem[Wayth et al.(2011)]{w11} Wayth, Randall B., Brisken, Walter F., Deller, Adam T., Majid, Walid A., Thompson, David R., Tingay, Steven J., \& Wagstaff, Kiri L. 2011, astro-ph/1104.4908, accepted to ApJ

 \bibitem[Welch et al.(2009)]{w09} Welch, J. et al., 2009, IEEEP, 97, 1438

 \bibitem[Wilson \& Weiler(1997)]{w97} Wilson, A. S. \& Weiler, K. W. 1997, ApJ, 475, 661


 \bibitem[von Korff et al.(2009)]{v09} von Korff, J., et al. 2009, ASPC, 420, 447

 \bibitem[Zarka et al.(1996)]{z96} Zarka, P., Farges, T., Ryabov, B. P., Abada-Simon, M., \& Denis, L. 1996, GeoRL, 23, 125

 \bibitem[Taylor(1974)]{t74} Taylor, J. H. 1974, A\&AS, 15, 367

 \end{thebibliography}
\end{document}